\documentclass[a4paper]{jpconf}
\usepackage{graphicx}

\usepackage{amssymb,amsmath}
\usepackage{cite}
\def\be{\begin{equation}}
\def\ee{\end{equation}}
\def\bea{\begin{eqnarray}}
\def\eea{\end{eqnarray}}
\def\beb{\begin{eqnarray*}}
\def\eeb{\end{eqnarray*}}

\begin{document}
\begin{flushright}
Preprint ESI 1971 (2007)
\end{flushright}

\title{Induced Gauge Theory on a Noncommutative Space\footnote{Invited talk given at the conference "Noncommutative Geometry and
Physics", Orsay, April 23-27, 2007}}

\author{Michael Wohlgenannt}

\address{Erwin Schr\"odinger International Institute for Mathematical Physics, 
Boltzmanngasse 9, A-1090 Vienna, Austria\\
and\\
Faculty of Physics, University of Vienna, Boltzmanngasse 5, A-1090 Vienna, Austria}

\ead{michael.wohlgenannt@univie.ac.at}

\begin{abstract}
We discuss the calculation of the 1-loop effective action on 
four dimensional, canonically deformed Euclidean space. The theory under consideration is
a scalar $\phi^4$ model with an additional oscillator potential. This model is known to be re
normalisable. Furthermore, we couple an exterior gauge field to the scalar field and 
extract the dynamics for the gauge field from the divergent terms of the 1-loop effective action 
using a matrix basis. This results in   
proposing an action for noncommutative gauge theory, which is a candidate 
for a renormalisable model.
\end{abstract}


\section{\label{introduction}Introduction}

This talk is based on a joint work with H.~Grosse. For more details see \cite{Grosse:2007dm}. The two 
dimensional case has been discussed in \cite{Grosse:2006hh}.

Feynman rules for Quantum Field Theory over noncommutative spaces reveal new structures. They stem
from the modification of space-time at small length scales. Planar contributions show the 
standard singularities which can be handled by the usual renormalisation procedure. The non-planar
one loop contributions are finite for generic momenta. However, they become logarithmically 
divergent at exceptional momenta. The usual UV divergences are then reflected in new singularities 
in the infrared, which is called UV/IR mixing. This spoils the usual renormalisation procedure: 
Inserting such loops to a higher order diagram generates singularities of any inverse power. 
In \cite{Grosse:2004yu}, H.~Grosse and R.~Wulkenhaar were able to give a solution 
of this problem for the special case of a scalar theory defined on the canonically 
deformed Euclidean space 
$\mathbb R^4_\theta$ with commutation relation for the coordinates:
$$
[x^\mu \stackrel{\star}{,} x^\nu] = i \theta^{\mu\nu},
$$
where $\theta^{ij}=-\theta^{ji}\in \mathbb{R}$,
and the $\star$-product is given by the Weyl-Moyal product
\be
f\star g \, (x) = e^{i/2\theta^{\mu\nu}\frac{\partial}{\partial x^\mu}\frac{\partial}{\partial y^\nu}}
f(x)g(y)\big|_{y\to x}.
\ee
For simplicity, we use the the following parametrisation of $\theta_{\mu\nu}$:
$$
(\theta_{\mu\nu}) = \left(
\begin{array}{cccc}
0 & \theta &  & \\
-\theta & 0 & & \\
&&0&\theta      \\
&&-\theta&0
\end{array}\right), \quad
(\theta^{-1}_{\mu\nu}) = \left(
\begin{array}{cccc}
0 & -1/\theta &&\\
1/\theta & 0&&  \\
&&0&-1/\theta   \\
&&1/\theta&0    
\end{array}
\right). 
$$
The UV/IR mixing contributions were taken into account through a modification of the
free Lagrangian by adding an oscillator term with parameter $\Omega$,
\bea
S_0 =  \int d^4 x &\left( \frac{1}{2} \phi \star 
[\tilde x_\nu, \,  [\tilde x^\nu, \phi]_\star]_\star 
+ \frac{\Omega^2}{2} 
\phi \star \{ \tilde x^\nu , \{ \tilde x_\nu ,\phi\}_\star \}_\star 
\right.
\nonumber
\\
& \left. + \frac{\mu^2}{2} \phi \star \phi 
+ \frac{\lambda}{4!} \phi\star \phi  \star \phi \star \phi\right)(x)\;,
\label{action1}
\eea
where $\tilde x_\nu = \theta^{-1}_{\nu\alpha}x^\alpha$ and 
$i\partial_\mu f = [\tilde x_\mu, f]_\star$.
The spectrum of the free Hamiltonian is modified. The harmonic oscillator term was
obtained as a result of the renormalisation proof. The model fulfills
the Langmann-Szabo duality \cite{Langmann:2002cc} relating short distance and long distance 
behaviour. There are indications that  a constructive procedure might be possible
and give a nontrivial $\phi^4$ model, which is currently under investigation \cite{Rivasseau:2005bh}.

In a different interesting approach, the UV/IR singularities are interpreted in terms of
an induced gravity action \cite{Steinacker:2007dq}.

In order to obtain the action for a gauge theory, which hopefully is renormalisable, we extract the 
divergent
terms of the heat kernel expansion. Such a procedure leads in the 
commutative case to a renormalisable gauge field action. 
We introduce the local, unitary gauge group $\mathcal G$ under which the scalar field $\phi$ 
transforms covariantly
like
\be
\label{i.1}
\phi \mapsto u^* \star \phi \star u, \,\, u\in \mathcal G.
\ee
The approach employed here makes use of two basic ideas. 
First, it is well known that the $\star$-multiplication of a coordinate - and also of a function, 
of 
course - with a field is not a covariant process. The product $x^\mu \star \phi$ will not transform
covariantly,
$$
x^\mu \star \phi \nrightarrow u^* \star x^\mu \star \phi \star u\;.
$$
Functions of the coordinates are not effected by the gauge group. Fields are taken to be
elements of a module \cite{Jurco:2000fs}. The introduction of covariant coordinates 
\be
\tilde X_\nu=\tilde x_\nu + A_\nu  
\ee
finds a remedy to this situation \cite{Madore:2000en}. The gauge field $A_\mu$ transforms such that
we have for the covariant coordinates:
\bea
\nonumber  
\tilde X_\mu  & \mapsto  & u^* \star \tilde X_\mu \star u\,;
\\
\label{i.2}
A_\mu & \mapsto & \mathrm{i} u^* \star \partial_\mu u  + u^* \star A_\mu
\star u
\; .
\eea
This leads to the definition of a gauge invariant model, which is the starting point
of our investigations. This model is given by the following action:  
\bea
S & = & \int d^4 x \left( 
\frac{1}{2} \phi \star [\tilde X_\nu,\, [\tilde X^\nu,\, \phi]_\star ]_\star 
+ \frac{\Omega^2}{2} 
\phi \star \{\tilde X^\nu , \{ \tilde X_\nu ,\phi\}_\star \}_\star 
\right.
\nonumber
\\
&&
+ \left. \frac{\mu^2}{2} \phi \star \phi 
+ \frac{\lambda}{4!} \phi\star \phi  \star \phi \star \phi\right)(x)\;.
\label{action}
\eea

Secondly, we apply the heat kernel formalism. The gauge field $A_\mu$ is an external, classical 
gauge
field coupled to $\phi$. In the following sections, we will explicitly 
calculate the divergent terms of the one-loop effective action. In the classical case, the divergent
terms determine the dynamics of the gauge field
\cite{Chamseddine:1996zu,Langmann:2001cv,Vassilevich:2003xt}.
There have already been attempts to generalise this approach to the non-commutative realm; for
non-commutative $\phi^4$ theory see \cite{Gayral:2004cs,Gayral:2004cu}.
First steps towards gauge kinetic models have been done in
\cite{Vassilevich:2003yz,Gayral:2004ww,Vassilevich:2005vk}. However, the results there are not 
completely comparable.
Our action contains an oscillator term
$$
\frac{\Omega^2}{2} 
\phi \star \{\tilde X^\nu , \{ \tilde X_\nu ,\phi\}_\star \}_\star\,.
$$
This term is crucial, it alters the free theory. 
Therefore, we expand around the free action $-\Delta + \Omega^2\tilde x^2$
rather than $-\Delta$.  As a consequence, the Seeley-de Witt coefficients cannot be used.

In the following sections, we describe our model and the employed method of extracting the 
singular contributions of the one-loop action in some detail. The results are summarised and discussed in the final Section.

\section{The Model}

The regularised one loop effective action for the model defined by the classical 
action~(\ref{action}) is given by
\be 
\Gamma^\epsilon_{1l}[\phi] = -\frac{1}{2} \int_\epsilon^\infty 
\frac{dt}{t} \,\mathrm{Tr}\left( e^{-t H} - e^{-t H^0} \right) \;. 
\label{Gamma-e} 
\ee
For the effective potential $H$ we have the expression 
\be
\frac \theta 2 \frac{\delta^2 S}{\delta \phi^2} \equiv H = H^0 + \frac \theta 2 V\,.
\ee
The field independent contributions are contained in the potential $H_0$, whereas $V$ 
involves linear and quadratic terms in the gauge and matter field.
The method is not manifestly gauge invariant, contributions from different orders need
to add up to a gauge invariant result. 

The effective action is calculated as a power series in the potential $V$. In order to do so
we employ the Duhamel expansion which is an iteration of the 
identity
\bea 
e^{-tH}-e^{-tH^0} &=& \int_0^t d\sigma \; \frac{d}{d \sigma} 
\left( e^{-\sigma H} e^{-(t-\sigma)H^0} \right) \nonumber
\\
&=& -\int_0^t d\sigma \; e^{-\sigma H} \,\frac{\theta}{2} V 
\,e^{-(t-\sigma)H^0} \;, 
\eea 
yielding 
\bea 
e^{-tH} & = & e^{-t H^0} - 
\frac{\theta}{2} \int_0^t d t_1 e^{-t_1 H^0} V e^{-(t-t_1) H^0} 
\nonumber
\\
&&+  \Big(\frac{\theta}{2}\Big)^2 \int_0^t d t_1 \int_0^{t_1} d t_2 
e^{-t_2 H^0} V e^{-(t_1-t_2) H^0} V e^{-(t-t_1) H^0} + \dots 
\eea
Therefore, we get for the 1-loop effective action the following formula:
\bea 
\nonumber
\label{duhamel}
\Gamma_{1l}^\epsilon & = & -\frac12 
\int_\epsilon^\infty \frac{dt}{t} \mathrm{Tr}\bigg( 
e^{-tH}-e^{-tH^0} \bigg) 
\\
& = &\frac{\theta}4 \int_\epsilon^\infty dt \textrm{ Tr }
    V e^{-tH^0} - \frac{\theta^2}8 \int_\epsilon^\infty \frac{dt}{t} \int_0^t dt'\, t'
    \textrm{ Tr } V e^{-t'H^0} V e^{-(t-t')H^0}\\
\nonumber
& & + \frac{\theta^3}{16} \int_\epsilon^\infty \frac{dt}{t} \int_0^t 
dt' \int_0^{t'} dt'' \, t''
    \textrm{ Tr } V e^{-t''H^0} V e^{-(t'-t'')H^0} V e^{-(t-t')H^0}\\
\nonumber
&& \hspace{-1.2cm}
-\frac{\theta^4}{32} \int_\epsilon^\infty \frac{dt}{t} \int_0^t 
dt' \int_0^{t'} dt'' 
    \int_0^{t''} dt''' \, t'''
    \textrm{ Tr } V e^{-t'''H^0} V e^{-(t''-t''')H^0} V e^{-(t'-t'')H^0} V e^{-(t-t')H^0}
\\
\nonumber
&& +\mathcal O(\theta^5)\, .
\eea

The calculations are performed in the matrix basis, where the star product is just a matrix
product:
$$
A^\nu(x) =\sum_{p,q \in \mathbb{N}^2} A^\nu_{pq} f_{pq}(x)\;, 
\phi(x) = \sum_{p,q \in \mathbb{N}^2} \phi_{pq} f_{pq}(x)
$$
and
\bea
f_{pq} \star f_{mn} & = & \delta_{qm} f_{pn},\\
f_0 \star f_0 & = & f_0\,.
\eea
This choice of basis simplifies the calculations. In the end, we will again represent the results 
in the $x$-basis. From the coordinates we can build two oscillators:
\be
a^{(1)} = \frac1{\sqrt{2}}(x^1+ix^2), \,\, a^{(2)} = \frac1{\sqrt{2}}(x^3+ix^4).
\ee
The ground state $f_0$ is a idempotent under
star multiplication and is given by a Gau\ss ian,
$$
f_0(x)  = 4 \, e^{-\frac1{\theta}\sum_i x_i^2}.
$$ 
All the other basis elements are obtained by
acting with creation and annihilation operators from the left and right, respectively:
\bea
f_{\stackrel{m^1}{m^2}\stackrel{n^1}{n^2}} & = & \alpha(n,m,\theta)
\, \bar {a}^{(2)\star m^2}\star \bar {a}^{(1)\star m^1} \star f_0 \star a^{(1)\star n^1} 
\star a^{(2)\star n^2},\\
\bar a^{(1)} \star f_{\stackrel{m^1}{m^2}\stackrel{n^1}{n^2}} & = & \sqrt{\theta(m^1+1)} 
f_{\stackrel{m^1+1}{m^2}\stackrel{n^1}{n^2}},\\
a^{(1)}\star f_{\stackrel{m^1}{m^2}\stackrel{n^1}{n^2}} & = & \sqrt{\theta m^1} 
f_{\stackrel{m^1-1}{m^2}\stackrel{n^1}{n^2}}\,.
\eea

In the next step, we have to apply the above method to the gauge invariant model~(\ref{action}).
After a suitable rescaling, all the operators depend, beside on $\theta$, only on the following three parameters:
\be
\rho = \frac{1-\Omega^2}{1+\Omega^2},\,
\tilde \epsilon = \epsilon (1+\Omega^2),\,
\tilde \mu^2 =  \frac{\mu^2 \theta}{1+\Omega^2}\,.
\ee
The part of the effective potential independent of the gauge field in the matrix basis is given by
\bea
\frac{H^0_{mn;kl}}{1+\Omega^2}
&=& \big(\frac{\tilde \mu^2}{2} {+} 
  (n^1{+}m^1{+}1) {+} (n^2{+}m^2{+}1) \big)
  \delta_{n^1k^1} \delta_{m^1l^1} \delta_{n^2k^2} \delta_{m^2l^2} \nonumber
  \\
&& \hspace{-2cm}
- \rho \big(\sqrt{k^1l^1}\,
  \delta_{n^1+1,k^1}\delta_{m^1+1,l^1 } 
+ \sqrt{m^1n^1}\, \delta_{n^1-1,k^1}
  \delta_{m^1-1,l^1} \big) \delta_{n^2k^2} \delta_{m^2l^2} \nonumber
  \\
&& \hspace{-2cm}
- \rho \big(\sqrt{k^2l^2}\,
  \delta_{n^2+1,k^2}\delta_{m^2+1,l^2 } 
+ \sqrt{m^2n^2}\, \delta_{n^2-1,k^2}
  \delta_{m^2-1,l^2} \big) \delta_{n^1k^1} \delta_{m^1l^1}\,.
\eea
For the field dependent potential $V$ we obtain
\bea
\frac {V_{kl;mn}} {(1+\Omega^2)} 
&=&\Big(
\frac{\lambda}{3!(1+\Omega^2)} \phi \star \phi
+ \big(\tilde X_\nu \star \tilde X^\nu -\tilde x^2 \big) \Big)_{lm} \delta_{nk}
\nonumber
\\
&&+ \Big(
\frac{\lambda}{3!(1+\Omega^2)} \phi \star \phi
+ \big(\tilde X_\nu \star \tilde X^\nu -\tilde x^2 \big) \Big)_{nk} \delta_{lm}
\nonumber
\\
&& + 
\Big(\frac{\lambda}{3!(1+\Omega^2)} \phi_{lm} \phi_{nk} 
- 2 \rho A_{\nu,lm} A^\nu_{nk} \Big) 
\nonumber
\\
&&+ \rho \mathrm{i}  \sqrt{\frac{2}{\theta}} \Big(
\sqrt{n^1} A^{(1+)}_{\stackrel{l^1}{l^2}\stackrel{m^1}{m^2}}
\delta_{\stackrel{n^1-1}{n^2}\stackrel{k^1}{k^2}}
- \sqrt{n^1+1}A^{(1-)}_{\stackrel{l^1}{l^2}\stackrel{m^1}{m^2}}
\delta_{\stackrel{n^1+1}{n^2}\stackrel{k^1}{k^2}}
\nonumber
\\
&&  \hspace*{7em}  
+ \sqrt{n^2} A^{(2+)}_{\stackrel{l^1}{l^2}\stackrel{m^1}{m^2}}
\delta_{\stackrel{n^1}{n^2-1}\stackrel{k^1}{k^2}}  
- \sqrt{n^2+1} A^{(2-)}_{\stackrel{l^1}{l^2}\stackrel{m^1}{m^2}}
\delta_{\stackrel{n^1}{n^2+1}\stackrel{k^1}{k^2}}\Big)
\nonumber
\\
&& 
- \rho \mathrm{i} \sqrt{\frac{2}{\theta}} \Big(
- \sqrt{m^1+1} A^{(1+)}_{\stackrel{n^1}{n^2}\stackrel{k^1}{k^2}} 
\delta_{\stackrel{l^1}{l^2}\stackrel{m^1+1}{m^2}}
+ \sqrt{m^1} A^{(1-)}_{\stackrel{n^1}{n^2}\stackrel{k^1}{k^2}} 
\delta_{\stackrel{l^1}{l^2}\stackrel{m^1-1}{m^2}}
\nonumber
\\
&& \hspace*{7em} - \sqrt{m^2+1} 
A^{(2+)}_{\stackrel{n^1}{n^2}\stackrel{k^1}{k^2}} 
\delta_{\stackrel{l^1}{l^2}\stackrel{m^1}{m^2+1}} + \sqrt{m^2} 
A^{(2-)}_{\stackrel{n^1}{n^2}\stackrel{k^1}{k^2}} 
\delta_{\stackrel{l^1}{l^2}\stackrel{m^1}{m^2-1}} \Big) 
\eea 
with the definitions
$$
A^{(1\pm)} = A^1\pm i A^2,\,\,
 A^{(2\pm)} = A^3\pm i A^4\,.
$$
The heat kernel $e^{-tH^0}$ of the Schr\"odinger operator
can be calculated from the propagator given in \cite{Grosse:2004yu}. In the matrix base
of the Moyal plane, it has the following representation:
\bea 
\left( e^{-tH^0}\right)_{mn;kl} & = & 
e^{-t(\mu^2\theta/2+ 4 \Omega)} \delta_{m+k,n+l} \prod_{i=1}^{2} 
K_{m^in^i;k^il^i}(t)\;,
\\
K_{m,m+\alpha;l+\alpha,l}(t) & = & \sum_{u=0}^{\textrm {min}(m,l)} 
\sqrt{\binom{m}{u}\binom{l}{u}
  \binom{\alpha+m}{m-u}\binom{\alpha+l}{l-u}} \\
\nonumber & &\times \, e^{2 \Omega t} \left( 
\frac{1-\Omega^2}{2\Omega}\sinh (2\Omega t) \right)^{m+l-2u} 
X_\Omega(t)^{\alpha+m+l+1} \;, 
\eea 
where 
\be
 X_\Omega(t)= \frac{4\Omega}{ (1+\Omega)^2e^{2\Omega t}-(1-\Omega)^2 e^{-2\Omega t}} \; .
\ee
The above expressions have to be inserted into the Duhamel expansion~(\ref{duhamel}).
Here, we are only interested in gauge theory. Therefore, we 
concentrate on the divergent terms involving only the gauge field and assume
$\lambda=0$.

\section{\label{calc}Some Remarks on the Calculation}

In order to extract the divergent contributions we employ the following method:
\begin{itemize}
\item First, expand the integrands of the Duhamel expansion~(\ref{duhamel})
for small auxiliary parameters $t,t',t'',\dots$.

\item Expand the infinite sums over indices occuring in the heat kernel but not in the gauge
field; divergences stem from these infinite sums. The other contractions are finite assuming that
$A$ is a traceclass operator.

\item Integrate over the auxiliary parameters.

\item Convert the results to x-space using
\be
\sum_m T_{mm} = \frac1{(2\pi\theta)^2}\int d^4x T(x)\,.
\ee
\end{itemize}
To first and second order in the potential $V$, the effective action contains both, logarithmic and 
quadratic divergences. To third and fourth order, only logarithmic ones occur. Higher powers in 
the potential are already finite. This can easily be seen from a power counting argument in the 
auxiliary parameters. Let us consider the contribution to the effective action of order $k$. Due
to Eq.~(\ref{duhamel}), there are $k$ auxiliary parameters. They for themselves produce a factor
$t^{k-1}$. The infinite sums over the integral kernels contribute inverse powers of $t$. For example,
we have in first order:
\bea
\sum_{n=0}^\infty K_{mn;nm}(t) & \sim & \sum_n X_\Omega(t)^n \sim \frac1{t}  +
    \mathcal O(t^0)\\
\sum_{n=0}^\infty \sqrt{n+1} K_{m+1,n+1;n,m}(t) & \sim &
\frac{\sqrt{m+1}}{t}  + \mathcal{O}(t^0) \,;
\eea
and in second order:
\bea
\nonumber
\sum_{n=0}^\infty K_{nm;mn}(t') K_{n+1,c;c,n+1}(t-t') & \sim &  \sum_n X_\Omega(t')^n X_\Omega(t-t')^n\sim
\frac1{t} + \mathcal O(t^0,t'^0)
\\
\sum_{n=0}^\infty \sqrt{n+1} K_{nm;m+1,n+1}(t') K_{n+1,c;c,n+1}(t-t')
& \sim &
\sqrt{m+1} \frac{t'}{t^2} + \mathcal O(t'^0, t^0)\,.
\eea
The potential $V$ may contribute in the worst case a factor $\sqrt{n}^k$ to the infinite sums of order $k$. Therefore, these
sums contribute a factor 
\bea
&&\hspace{-1.5cm}
\sum_n n^{k/2} X_\Omega(t^{(k)})^n X_\Omega(t^{(k-1)}-t^{(k)})^n \dots X_\Omega(t)^n 
\\
&&
\times \sum_m  X_\Omega(t^{(k)})^m X_\Omega(t^{(k-1)}-t^{(k)})^m \dots X_\Omega(t)^m
\sim 
\left(\frac1{t}\right)^{\lfloor k/2\rfloor+2},
\eea
where $\lfloor l\rfloor$ is the greatest integer function (see e.g. Mathematica for an exact definition).
Hence, the contribution to order $k$ is given by
\be
\left( \frac1{t} \right)^{ \lfloor k/2\rfloor + 3 - k }.
\ee
For $k=1$, the exponent is $2$, which means that quadratic divergences occur. In the case of $k=5$, the exponent is $0$ and the integration 
yields a finite result.

Details of the calculations are provided in \cite{Grosse:2007dm}.

\section{\label{results}Results and Conclusions}

Let us summarise the results. In the selfdual case, $\Omega=1$ the divergent contributions are of an especially simple form.
The matrix base expressions for the effective potential and the heat kernel simplify a lot. The effective action describes 
a pure matrix model. The one-loop effective action is given by
\bea
\label{result1}
\Gamma^\epsilon_{1l} & = &
\frac1{16\pi^2} \int d^4 x\, \Bigg(  \frac1{\epsilon\theta}
(\tilde X_\nu \star \tilde X^\nu -\tilde x^2)\\
\nonumber && \hspace{-1cm}
+ \bigg( \frac{\mu^2}2 (\tilde X_\nu \star \tilde X^\nu -\tilde x^2)
+ \frac12 \left( (\tilde X_\mu\star \tilde X^\mu)\star (\tilde X_\nu\star \tilde X^\nu)- (\tilde
x^2)^2 \right) \bigg) \ln \epsilon \Bigg) \,.
\eea
In this case, we propose the logarithmically divergent part as action for the gauge field:
\be
S = \frac1{16\pi^2} \int d^4 x\bigg( \frac{\mu^2}2 (\tilde X_\nu \star \tilde X^\nu -\tilde x^2)
+ \frac12 \left( (\tilde X_\mu\star \tilde X^\mu)\star (\tilde X_\nu\star \tilde X^\nu)- (\tilde
x^2)^2 \right) \bigg)\,.
\ee

In the case $\Omega\ne 0$, we obtain much more structure and a dynamics:
\begin{eqnarray}
\label{result2}
\Gamma_{1l}^\epsilon & = & \frac{1}{192\pi^2}  \int d^4x\, \Bigg\{
\frac{24}{\tilde \epsilon \, \theta} (1-\rho^2)(\tilde X_\nu \star \tilde X^\nu -\tilde x^2)\\
\nonumber
&&
+ \ln\epsilon \bigg(
\frac{12}{\theta} (1-\rho^2) (\tilde \mu^2-\rho^2)(\tilde X_\nu \star \tilde X^\nu -\tilde x^2) 
\\
\nonumber
&& \hspace{1.3cm}
+ 6(1-\rho^2)^2 \big( (\tilde X_\mu\star \tilde X^\mu)^{\star 2}-(\tilde x^2)^2 \big)
- \rho^4  F_{\mu\nu} F^{\mu\nu}
\bigg)
\Bigg\}\,,
\end{eqnarray}
where $F_{\mu\nu} = -i[\tilde x_\mu, A_\nu]_\star + i[\tilde x_\nu, A_\mu]_\star 
    -i [A_\mu,A_\nu]_\star$ .
Again, we propose the logarithmically divergent part as an action describing the dynamics of the gauge field,
\bea
S & = & \frac{1}{192\pi^2}  \int d^4x\, \Bigg\{
\frac{12}{\theta} (1-\rho^2) (\tilde \mu^2-\rho^2)(\tilde X_\nu \star \tilde X^\nu -\tilde x^2) 
\\
\nonumber
&& \hspace{1.3cm}
+ 6(1-\rho^2)^2 \big( (\tilde X_\mu\star \tilde X^\mu)^{\star 2}-(\tilde x^2)^2 \big)
- \rho^4  F_{\mu\nu} F^{\mu\nu} \Bigg\}\,.
\eea
Both, the linear in  $\epsilon$ and the logarithmic
in $\epsilon$ divergent term of the one-loop effective action turn out to be gauge invariant. 
The logarithmically divergent part is an interesting
candidate for a renormalisable gauge interaction. 
The sign of the term quadratic in the covariant coordinates may change depending on whether $\tilde\mu^2\lessgtr \rho^2$.
This reflects the structure of a phase transition.  The case $\Omega=1$ ($\rho=0$) is of course of 
particular interest. One obtains a pure matrix model.
In the limit $\Omega\to 0$, we obtain just the standard deformed Yang-Mills action.
Furthermore, the action ~(\ref{result2}) allows to study the limit $\theta\to \infty$.

In addition, we will attempt to study the perturbative quantisation.
One of the problems of quantising action (\ref{result2}) is connected to the tadpole contribution,
which is non-vanishing and hard to eliminate. The Orsay group also considered the 1-loop effective action in the
case $\Omega\ne 0$. They calculated the divergent contributions in x-space by evaluating Feynman diagrams and  
arrived at the same result \cite{deGoursac:2007gq, deGoursac:2007qi}.

Solutions of the equations of motion for similar models have already been considered in \cite{deGoursac:2007uv,Grosse:2007jy}.
An appropriate rescaling of the covariant coordinates $\tilde X_\alpha \to \frac{\sqrt{2\sqrt{3}}}{\sqrt{\theta}} \tilde X_\alpha$ and the 
identification $\tau \equiv - \sqrt{3}\,\frac{1-\rho^2}{\rho^2}$ leads to the equations of motion
\bea
\label{eom}
D_\nu F^{\sigma\nu} =\tau \tilde X^\sigma + \tau^2 \{ \tilde X^\sigma, \tilde X_\nu \star \tilde X^\nu\}_\star\,,
\eea
where we have assumed for simplicity $\tilde \mu = 0$ and used
$$
D_\nu F^{\sigma\nu} = -i [\tilde X_\nu, -i [\tilde X^\sigma,\tilde X^\nu]_\star + \theta^{-1\,\mu\nu}]_\star
= - [\tilde X_\nu, [\tilde X^\sigma,\tilde X^\nu]_\star ]_\star\,.
$$
In \cite{Grosse:2007jy}, the matter fields have been included in order to find some solutions. However, the gauge part (\ref{eom})
alone also exhibits a number of solutions which are currently under investigation.

For noncommutative $U(1)$ gauge theory a similar model has been discussed in \cite{Blaschke:2007vc}. 
This model includes an oscillator potential for the gauge fields, $\tilde x^2A^2$. Other terms 
occuring here are missing. Hence, the considered action is not gauge invariant, but a BRST invariance could be established.
These terms may nevertheless come into the game through one loop corrections. 

\section*{References}

\bibliographystyle{../../latex-styles/utphys}
\bibliography{../../tuw}

\providecommand{\href}[2]{#2}\begingroup\raggedright\begin{thebibliography}{10}

\bibitem{Grosse:2007dm}
H.~Grosse and M.~Wohlgenannt, ``Induced gauge theory on a noncommutative
  space,'' {\em Eur. Phys. J.} {\bf C52} (2007) 435--450,
\href{http://www.arXiv.org/abs/hep-th/0703169}{{\tt hep-th/0703169}}.

\bibitem{Grosse:2006hh}
H.~Grosse and M.~Wohlgenannt, ``Noncommutative {QFT} and renormalization,''
  {\em J. Phys. Conf. Ser.} {\bf 53} (2006) 764--792,
\href{http://www.arXiv.org/abs/hep-th/0607208}{{\tt hep-th/0607208}}.

\bibitem{Grosse:2004yu}
H.~Grosse and R.~Wulkenhaar, ``Renormalisation of {$\phi^4$} theory on
  noncommutative {$\mathbb R^4$} in the matrix base,'' {\em Commun. Math.
  Phys.} {\bf 256} (2005) 305--374,
\href{http://www.arXiv.org/abs/hep-th/0401128}{{\tt hep-th/0401128}}.

\bibitem{Langmann:2002cc}
E.~Langmann and R.~J. Szabo, ``Duality in scalar field theory on noncommutative
  phase spaces,'' {\em Phys. Lett.} {\bf B533} (2002) 168--177,
\href{http://www.arXiv.org/abs/hep-th/0202039}{{\tt hep-th/0202039}}.

\bibitem{Rivasseau:2005bh}
V.~Rivasseau, F.~Vignes-Tourneret, and R.~Wulkenhaar, ``Renormalization of
  noncommutative $\phi^4$-theory by multi- scale analysis,'' {\em Commun. Math.
  Phys.} {\bf 262} (2006) 565--594,
\href{http://www.arXiv.org/abs/hep-th/0501036}{{\tt hep-th/0501036}}.

\bibitem{Steinacker:2007dq}
H.~Steinacker, ``Emergent gravity from noncommutative gauge theory,''
\href{http://www.arXiv.org/abs/arXiv:0708.2426 [hep-th]}{{\tt arXiv:0708.2426
  [hep-th]}}.

\bibitem{Jurco:2000fs}
B.~Jur\v{c}o, P.~Schupp, and J.~Wess, ``Noncommutative gauge theory for
  {Poisson} manifolds,'' {\em Nucl. Phys.} {\bf B584} (2000) 784--794,
\href{http://arXiv.org/abs/hep-th/0005005}{{\tt hep-th/0005005}}.

\bibitem{Madore:2000en}
J.~Madore, S.~Schraml, P.~Schupp, and J.~Wess, ``Gauge theory on noncommutative
  spaces,'' {\em Eur. Phys. J.} {\bf C16} (2000) 161--167,
\href{http://www.arXiv.org/abs/hep-th/0001203}{{\tt hep-th/0001203}}.

\bibitem{Chamseddine:1996zu}
A.~H. Chamseddine and A.~Connes, ``The spectral action principle,'' {\em
  Commun. Math. Phys.} {\bf 186} (1997) 731--750,
\href{http://www.arXiv.org/abs/hep-th/9606001}{{\tt hep-th/9606001}}.

\bibitem{Langmann:2001cv}
E.~Langmann, ``Generalized {Yang-Mills actions from Dirac} operator
  determinants,'' {\em J. Math. Phys.} {\bf 42} (2001) 5238--5256,
\href{http://www.arXiv.org/abs/math-ph/0104011}{{\tt math-ph/0104011}}.

\bibitem{Vassilevich:2003xt}
D.~V. Vassilevich, ``Heat kernel expansion: User's manual,'' {\em Phys. Rept.}
  {\bf 388} (2003) 279--360,
\href{http://www.arXiv.org/abs/hep-th/0306138}{{\tt hep-th/0306138}}.

\bibitem{Gayral:2004cs}
V.~Gayral, ``Heat-kernel approach to {UV/IR} mixing on isospectral deformation
  manifolds,'' {\em Annales Henri Poincare} {\bf 6} (2005) 991--1023,
\href{http://www.arXiv.org/abs/hep-th/0412233}{{\tt hep-th/0412233}}.

\bibitem{Gayral:2004cu}
V.~Gayral, J.~M. Gracia-Bondia, and F.~R. Ruiz, ``Trouble with space-like
  noncommutative field theory,'' {\em Phys. Lett.} {\bf B610} (2005) 141--146,
\href{http://www.arXiv.org/abs/hep-th/0412235}{{\tt hep-th/0412235}}.

\bibitem{Vassilevich:2003yz}
D.~V. Vassilevich, ``Non-commutative heat kernel,'' {\em Lett. Math. Phys.}
  {\bf 67} (2004) 185--194,
\href{http://www.arXiv.org/abs/hep-th/0310144}{{\tt hep-th/0310144}}.

\bibitem{Gayral:2004ww}
V.~Gayral and B.~Iochum, ``The spectral action for {M}oyal planes,'' {\em J.
  Math. Phys.} {\bf 46} (2005) 043503,
\href{http://www.arXiv.org/abs/hep-th/0402147}{{\tt hep-th/0402147}}.

\bibitem{Vassilevich:2005vk}
D.~V. Vassilevich, ``Heat kernel, effective action and anomalies in
  noncommutative theories,'' {\em JHEP} {\bf 08} (2005) 085,
\href{http://www.arXiv.org/abs/hep-th/0507123}{{\tt hep-th/0507123}}.

\bibitem{deGoursac:2007gq}
A.~de~Goursac, J.-C. Wallet, and R.~Wulkenhaar, ``Noncommutative induced gauge
  theory,''
\href{http://www.arXiv.org/abs/hep-th/0703075}{{\tt hep-th/0703075}}.

\bibitem{deGoursac:2007qi}
A.~de~Goursac, ``On the effective action of noncommutative yang-mills theory,''
\href{http://www.arXiv.org/abs/arXiv:0710.1162 [hep-th]}{{\tt arXiv:0710.1162
  [hep-th]}}.

\bibitem{deGoursac:2007uv}
A.~de~Goursac, A.~Tanasa, and J.~C. Wallet, ``Vacuum configurations for
  renormalizable non-commutative scalar models,''
\href{http://www.arXiv.org/abs/arXiv:0709.3950 [hep-th]}{{\tt arXiv:0709.3950
  [hep-th]}}.

\bibitem{Grosse:2007jy}
H.~Grosse and R.~Wulkenhaar, ``8d-spectral triple on 4d-moyal space and the
  vacuum of noncommutative gauge theory,''
\href{http://www.arXiv.org/abs/arXiv:0709.0095 [hep-th]}{{\tt arXiv:0709.0095
  [hep-th]}}.

\bibitem{Blaschke:2007vc}
D.~N. Blaschke, H.~Grosse, and M.~Schweda, ``Non-commutative u(1) gauge theory
  on r**4 with oscillator term,''
\href{http://www.arXiv.org/abs/arXiv:0705.4205 [hep-th]}{{\tt arXiv:0705.4205
  [hep-th]}}.

\end{thebibliography}\endgroup

\end{document}